\documentclass[aps,pre,preprint,superscriptaddress,floatfix]{revtex4-1}
\usepackage{bm}
\usepackage{graphicx,color}
\usepackage{graphicx}
\usepackage{dcolumn}
\usepackage{times}

\usepackage{amsmath,amssymb}

\begin{document}

\title{Scale-free chaos in the 2D harmonically confined Vicsek model} 
\author{R. Gonz\'alez-Albaladejo}
\affiliation{Departamento de Matem\'atica Aplicada, Universidad Complutense de Madrid, 28040 Madrid, Spain}
\affiliation{Gregorio Mill\'an Institute for Fluid Dynamics, Nanoscience and Industrial Mathematics, Universidad Carlos III de Madrid, 28911 Legan\'{e}s, Spain}
\author{L. L. Bonilla$^*$}
\affiliation{Gregorio Mill\'an Institute for Fluid Dynamics, Nanoscience and Industrial Mathematics, Universidad Carlos III de Madrid, 28911 Legan\'{e}s, Spain}
\affiliation{Department of Mathematics, Universidad Carlos III de Madrid, 28911 Legan\'{e}s, Spain. 
$^*$Corresponding author. E-mail: bonilla@ing.uc3m.es}
\date{\today}
\begin{abstract}
Animal motion and flocking are ubiquitous nonequilibrium phenomena that are often studied within active matter. In examples such as insect swarms, macroscopic quantities exhibit power laws with measurable critical exponents and ideas from phase transitions and statistical mechanics have been explored to explain them. The widely used Vicsek model with periodic boundary conditions has an ordering phase transition but the corresponding homogeneous ordered or disordered phases are different from observations of natural swarms. If a harmonic potential (instead of a periodic box) is used to confine particles, the  numerical simulations of the Vicsek model display periodic, quasiperiodic and chaotic attractors. The latter are scale free on critical curves that produce power laws and critical exponents. Here we investigate the scale-free-chaos phase transition in two space dimensions. We show that the shape of the chaotic swarm on the critical curve reflects the split between core and vapor of insects observed in midge swarms and that the dynamic correlation function collapses only for a finite interval of small scaled times, as also observed. We explain the algorithms used to calculate the largest Lyapunov exponents, the static and dynamic critical exponents and compare them to those of the three-dimensional model.
\end{abstract}

\maketitle

\section{Introduction}\label{sec:1}
Experiments on the social behavior of animals in the laboratory may give rise to unreproducible results due to imposing artificial tasks on the animals and subsets of animals behaving differently  \cite{pus22}. Experiments in natural environments may allow for observing the emergence of social behavior free from artificial laboratory constraints. Examples include flocks of birds \cite{bal08} and sheep \cite{gin15}, fish schools \cite{her11}, marching locusts \cite{buh06}, swarms of midges \cite{att14plos} and hordes of rodents \cite{kon19,che23}. In these systems, collective behavior results from the dynamical interaction between individuals often producing power laws, which poses the question whether biological systems are in the critical region of a phase transition \cite{mor11,bia12}. 

Many works have tried to apply ideas from phase transitions in statistical mechanics (scale-free behavior, finite size scaling, renormalization group, critical exponents, universality \cite{hua87,wil83,ami05,hoh77}) to the emergent collective behavior in animals and to compare them to experiments \cite{cav18}.  Many theoretical studies have dealt with the Vicsek model (VM) of particles moving at a constant speed in a box with periodic boundary conditions and changing their velocity at discrete times by selecting the mean average velocity of all particles in their neighborhood plus an alignment noise \cite{vic95,vic12,cha20}. This model presents an ordering phase transition when the alignment noise falls below a critical value \cite{cha20}, which reminds of an equilibrium second-order phase transition between disordered and ordered homogeneous phases at the critical temperature \cite{hua87}. For animal collectivities, the periodic boundary conditions are artificial and it has been argued that placing the particles in a confining harmonic potential keeps the cohesion of the flock \cite{oku86,kel13,gor16,cav17}. 

 In recent works, we have numerically simulated the three-dimensional (3D) harmonically confined Vicsek model (HCVM) and discovered a phase transition between phases within chaos \cite{gon23}. We have also analyzed the mean field HCVM near the scale-free-chaos phase transition and found that its static critical exponents are the same as in the Landau theory of equilibrium phase transitions \cite{gon23mf}. The critical exponents obtained from our numerical simulations \cite{gon23,gon23arxiv} are close to those measured in natural midge swarms \cite{cav17,att14,cav23}. In swarms, midges acoustically interact when their distances are sufficiently small \cite{att14}. The distribution of speeds is peaked about some value and exhibits heavy tails for large swarms (perhaps due to the formation of clusters) \cite{kel13}. The statistics of accelerations of individual midges in a swarm is consistent with
postulating a linear spring force (therefore a harmonic potential) that binds insects together \cite{kel13}. Laboratory experiments have shown that the swarm consists of a condensed core and a vapor of insects that leave or enter it \cite{sin17}. Swarms of midges form over specific darker spots
on the ground (wet areas, cow dung, man-made objects, etc) called markers \cite{dow55}. This empirical fact precludes states of the swarm that are invariant under space translations, such as the ordered and disordered homogeneous phases that are conspicuous in theories of active matter \cite{cha20,cav23}.

Other animals in a flock, such as starlings \cite{bal08}, define neighbors topologically not metrically. Bird rotations propagate swiftly as linear waves \cite{cav18}. Metric-free models may incorporate a distributed motional bias \cite{lew17}, visual and auditory sensing are compared in \cite{roy19}, the influence of time delay is studied in \cite{gei22}, the influence of metric and topological interactions on flocking is studied in \cite{kum21} and \cite{rey17} considers a swarming model based on effective velocity-dependent gravity. 
 
 Here we consider the 2D HCVM, which may model large vertical insect swarms \cite{att14}. As in the 3D case, the 2D HCVM exhibits scale-free chaos on a curve in the parameter space of alignment noise and spring constant (confinement parameter) for finitely many particles. This curve separates chaotic states in which the swarm is split into several subgroups from chaotic single-cluster states \cite{gon23}. On the critical curve, the swarm size is proportional to the correlation length, which is the only length scale entering power laws for macroscopic quantities. Thus, the swarm is scale free on the critical curve. Using the finite size and dynamical scaling hypotheses \cite{ami05,hoh77}, we can extrapolate power laws obtained for finitely many particles to the phase transition comprising infinitely many particles \cite{att14,cav17,cav18,cav23}, which we call scale-free-chaos phase transition \cite{gon23}. For infinitely many particles, the confinement parameter and the largest Lyapunov exponent are both zero \cite{gon23}. In 3D, there are other critical curves that coalesce at the same rate to the line of zero confinement as the number of particles $N\to\infty$. For finite $N$, these critical curves form an extended criticality region, which may describe data from natural swarms \cite{gon23arxiv}. We calculate the static and dynamic critical exponents and show their relation to the 3D ones. We show that the origin of the parameter space (zero confinement and zero noise) is an organizing center that helps understanding the scale-free-chaos phase transition. In particular, it is possible to calculate one static critical exponent by varying the noise at fixed $N$ instead of the traditional approach of varying $N$ for confinement on the critical curve at fixed noise strength \cite{gon23arxiv}.
 
 The rest of the paper is organized as follows. Section~\ref{sec2} describes the HCVM and the deterministic attractors that appear at different values of the confinement strength for zero noise in numerical simulations. We find periodic, quasiperiodic and chaotic attractors. To distinguish the latter, we chiefly use the Benettin algorithm (BA) \cite{ben80} to compute regions in parameter space where the largest Lyapunov exponent (LLE) is positive. The alignment noise of the HCVM changes the attractors as explained in Section~\ref{sec3}. Section~\ref{sec4} describes the extended criticality region of the scale-free-chaos phase transition and the associated power laws with critical exponents determined from our numerical simulations. For this purpose, we use the critical curve separating single and multicluster chaos: on it, we fix the strength of the noise and vary the number of particles $N$. In Section \ref{sec5}, we determine critical exponents by fixing $N$ and varying the noise strength on an interval of small values. Section \ref{sec6} discusses our results and contains our conclusions.
  
 \section{HCVM and its deterministic attractors}\label{sec2}
In 2D, the HCVM consists of $N$ particles with positions $\mathbf{x}_j(t)=(x_j(t),y_j(t))$ and velocities $\mathbf{v}_j(t)=v\,(\cos\theta_j(t),\sin\theta_j(t))$, $j=1,\ldots,N$, that are updated at discrete times $m\Delta t$, $m=0,1,\ldots$, according to the rule, 
\begin{subequations}\label{eq1}
\begin{eqnarray}
&&\mathbf{x}_j(t\!+\!\Delta t)=\mathbf{x}_j(t)+ \Delta t\,\mathbf{v}_j(t+\Delta t),\quad z_j=x_j+iy_j,  \label{eq1a}\\
&&\theta_j(t\!+\!\Delta t)\!=\!\mbox{Arg}\!\left(\sum_{|\mathbf{x}_k\!-\mathbf{x}_j|<r_1\!R_0}\!\!e^{i\theta_j\!(t)}\!-\!\frac{\beta_0}{v}z_j(t)\!\right)\!\!+\!\xi_j(t)\!.\quad  \label{eq1b}
\end{eqnarray}
In Eq.~\eqref{eq1b}, we sum over all particles (including $j$) that, at time $t$, are inside a circle of influence with radius $r_1R_0$ centered at $\mathbf{x}_j(t)$; $r_1$ is the time-averaged nearest-neighbor distance within the swarm and $R_0$ is the dimensionless radius. At each time, $\xi_j(t)$ is a random number chosen with equal probability in the interval $(-\eta/2,\eta/2)$.

We nondimensionalize the model using data from the observations of natural midges reported in the supplementary material of Ref.~\onlinecite{cav17}. We measure times in units of $\Delta t=0.24$ s, lengths in units of the time-averaged nearest-neighbor distance of the $20120910\_A1$ swarm  in Table I \cite{cav17}, which is $r_1=4.68$ cm, and velocities in units of $r_1/\Delta t$, whereas $v=0.195$ m/s. The nondimensional version of Eq.~(\ref{eq1}) has $\Delta t=1$, $r_1=1$, speed $v_0$ and confinement parameter $\beta$ given by
\begin{eqnarray}
v_0=v\, \frac{\Delta t}{r_1}, \quad \beta=\beta_0 \Delta t. \label{eq1c}
\end{eqnarray} \end{subequations}
For the example we have selected, $v_0=1$, whereas other entries in the same table produce order-one values of $v_0$ with average 0.53. For these values, the HCVM has the same behavior as described here. Thus, our HCVM describing midge swarms is far from the continuum limit $v_0\ll 1$. Cavagna {\em et al} consider a much smaller speed for their periodic VM, $v_0=0.05$, closer to the continuum limit where derivatives replace finite differences \cite{cav17}. Thus, the nondimensional equations of the HCVM are
\begin{subequations}\label{eq2}
\begin{eqnarray}
&&\mathbf{x}_j(t\!+1)=\mathbf{x}_j(t)+ \mathbf{v}_j(t+1),\,\, \mathbf{v}_j=v_0(\cos\theta_j,\sin\theta_j),\quad  \label{eq2a}\\
&&\theta_j(t\!+1)\!=\!\mbox{Arg}\!\left(\sum_{|\mathbf{x}_k\!-\mathbf{x}_j|<R_0}\!\!e^{i\theta_j\!(t)}\!-\frac{\beta}{v_0} z_j(t)\!\right)\!\!+\!\xi_j(t).  \label{eq2b}
\end{eqnarray}
Eq.~\eqref{eq2b} can be written equivalently as
\begin{eqnarray}
&& \mathbf{v}_j(t+1)=v_0  \mathcal{R}_\eta \Theta\!\left(\sum_{|\mathbf{x}_j-\mathbf{x}_i|<R_0}\mathbf{v}_j(t)-\beta\mathbf{x}_i(t)\right)\!, \label{eq2c}\\
&&\mathcal{R}_\eta= \begin{pmatrix} 
\cos\xi& -\sin\xi\\ 
\sin\xi&\cos\xi
\end{pmatrix}\!,\quad-\frac{\eta}{2}\leq\xi\leq\frac{\eta}{2}.\label{eq2d}
\end{eqnarray}
\end{subequations}
Here $\Theta(\mathbf{x})=\mathbf{x}/|\mathbf{x}|$ and $\xi$ is a random number selected with equal probability on the specified interval of width $\eta$. 

The HCVM has chaotic attractors among its solutions for appropriate values of the parameters. We identify these attractors by calculating the largest Lyapunov exponent (LLE), which is positive for chaos. To this effect, we use the Benettin algorithm (BA) \cite{ben80}. We have to simultaneously solve Equations~\eqref{eq2} and the linearized system of equations 
\begin{subequations}\label{eq3}
\begin{eqnarray}
\delta\mathbf{\tilde{x}}_i(t+1)\!&=&\! \delta\mathbf{\tilde{x}}_i(t)+\delta\mathbf{\tilde{v}}_i(t+1),  \quad i=1,\ldots, N,  \label{eq3a}\\
\delta\mathbf{\tilde{v}}_i(t+1)\!&=&\! v_0\mathcal{R}_\eta\!\left(\mathbb{I}_2-\frac{[\sum_{|\mathbf{x}_j-\mathbf{x}_i|<R_0}\mathbf{v}_j(t)-\beta\mathbf{x}_i(t)]^T[\sum_{|\mathbf{x}_j-\mathbf{x}_i|<R_0}\mathbf{v}_j(t)-\beta\mathbf{x}_i(t)]}{|\sum_{|\mathbf{x}_j-\mathbf{x}_i|<R_0}\mathbf{v}_j(t)-\beta\mathbf{x}_i(t)|^2}\right)\nonumber\\
\!&\cdot &\!\frac{\sum_{|\mathbf{x}_j-\mathbf{x}_i|<R_0}\delta\mathbf{\tilde{v}}_j(t)-\beta\delta\mathbf{\tilde{x}}_i(t)}{|\sum_{|\mathbf{x}_j-\mathbf{x}_i|<R_0}\mathbf{v}_j(t)-\beta\mathbf{x}_i(t)|} ,  \label{eq3b}
\end{eqnarray}\end{subequations}
in such a way that the random realizations $\mathcal{R}_\eta$ are exactly the same for Equations~\eqref{eq2} and \eqref{eq3}. Here $\mathbb{I}_2$ is the 2D identity matrix. The initial conditions for the disturbances, $\delta\mathbf{\tilde{x}}_i(0)$ and $\delta\mathbf{\tilde{v}}_i(0)$, can be randomly selected so that the overall length of the vector $\delta\bm{\chi}=(\delta\mathbf{\tilde{x}}_1,\ldots,\delta\mathbf{\tilde{x}}_N,\delta\mathbf{\tilde{v}}_1,\ldots,\delta\mathbf{\tilde{v}}_N)$ equals 1. After each time step $t$, the vector $\delta\bm{\chi}(t)$ has length $\alpha_t$. At that time, we renormalize $\delta\bm{\chi}(t)$ to $\hat{\bm{\chi}}(t)=\delta\bm{\chi}(t)/\alpha_t$ and use this value as initial condition to calculate $\delta\bm{\chi}(t+1)$. With all the values $\alpha_t$ and for sufficiently large $l$, we calculate the Lyapunov exponent as
\begin{eqnarray}
\lambda_1= \frac{1}{l}\sum_{t=1}^l\ln\alpha_t, \label{eq4}\quad
 \alpha_t=|\delta\bm{\chi}(t)|=|(\delta\mathbf{\tilde{x}}_1(t),\ldots,\delta\mathbf{\tilde{x}}_N(t),\delta\mathbf{\tilde{v}}_1(t),\ldots,\delta\mathbf{\tilde{v}}_N(t))|.
\end{eqnarray}
See Figures 17 and 18 of Ref.~\cite{gon23} for convergence of the BA.

\begin{center}
\begin{figure}[ht]
\begin{center}
\includegraphics[clip,width=0.97\linewidth]{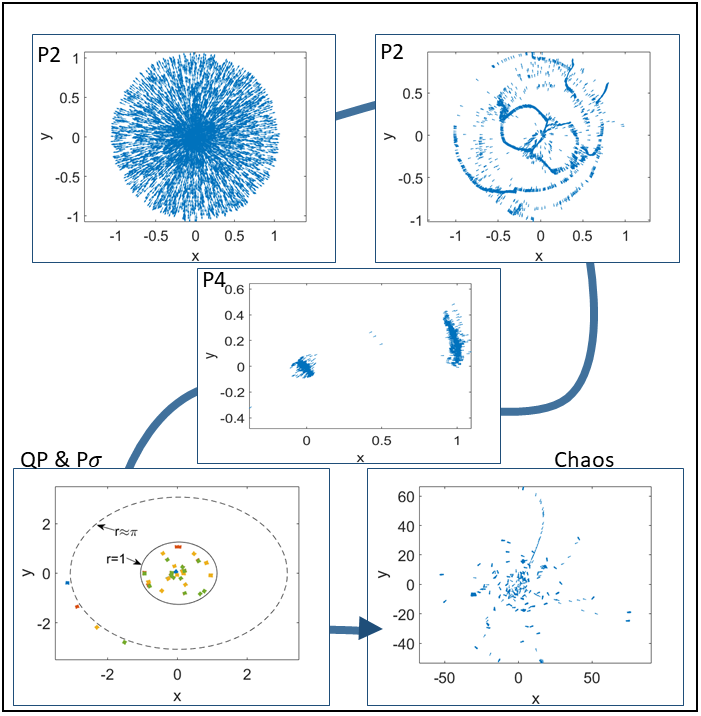}\\
\end{center}
\caption{Visual patterns for $N = 5000, R_0 = 0.472, \eta = 0$ and decreasing values of $\beta$. Period-2 solutions occur at $\beta=10^9$ and at $\beta=2\times 10^5$: the particles remain inside the unit circle forming either connected patterns or patterns with large holes (reminding of the vibration mode of a plate). At $\beta=10^4$, there are two subgroups within the unit circle that exchange their positions in a period-4 solution. At $\beta=500$, there are large period solutions with subgroups of particles inside the unit circle and others oscillating outside, forming orbits reminiscent of the Bohr atomic model (the different colors correspond to different times). At $\beta=0.1$, the pattern corresponds to a chaotic attractor displaying many more subgroups (of smaller density) and also single particles.  \label{fig1}}
\end{figure}
\end{center}

Figure \ref{fig1} shows different attractors for $N=5000$ in the deterministic case $\eta=0$. Initially, the particles are randomly placed within a sphere with unit radius and the particle velocities are pointing outwards. As $\beta$ decreases from very large values, we observe, from top to bottom and from the left to the right, period two (P2) solutions, period-four (P4) solutions, quasiperiodic (QP), large period solutions and chaotic solutions. At  $\beta= 10^9$, the particles remain inside the unit circle forming a symmetric pulsating pattern that repeats itself every two iterations. P2 solutions also occur for $\beta =2\times 10^5$ but the particles form a pattern with unoccupied spaces within the unit circle. At $\beta = 10^4$, the period has duplicated, the particles form two subgroups and pass from one to the other, repeating their positions every four iterations. At $\beta=500$, there are large period solutions with subgroups of particles inside the unit circle and others oscillating outside, forming orbits reminiscent of the Bohr atomic model (the different colors correspond to different times). Close to these solutions, there are quasiperiodic ones. At $\beta=0.1$, quasiperiodicity has evolved to chaos, as shown in the last panel of Fig.~\ref{fig1}.

\begin{center}
\begin{figure}[ht]
\begin{center}
\includegraphics[clip,width=0.97\linewidth]{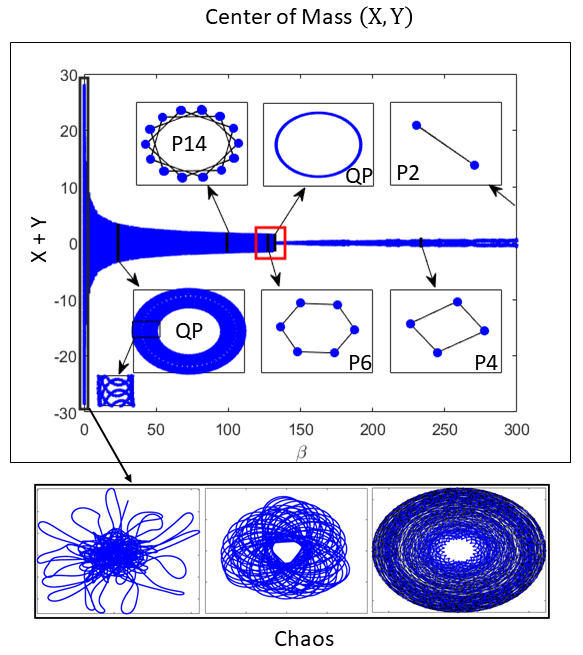}\\
\end{center}
\caption{Bifurcation diagram of the center of mass coordinates $X+Y$ versus $\beta$ for $N=128$. At different values of confinement there are P2, P4, P6, quasi-periodic solutions interspersed with  larger-period solutions (e.g., P14), and different chaotic attractors for smaller $\beta$.   \label{fig2}}
\end{figure}
\end{center}

Fig.~\ref{fig2} shows the bifurcation diagram of the center of mass (CM) coordinates $X+Y$ versus $\beta$ for $N=128$. Here $\mathbf{X}(t)= N^{-1}\sum_{j=1}^N \mathbf{x}_j(t)$. At decreasing values of $\beta$, there are P2, P4, P6, quasi-periodic solutions interspersed with larger-period solutions (e.g., P14), and different chaotic attractors for smaller $\beta$. The CM trajectories illustrate the different types of attractors in the HCVM.

\section{Noisy attractors of the HCVM}\label{sec3}
In this section, we describe the attractors of the noisy HCVM and characterize the regions of deterministic and noisy chaos in parameter space. To this purpose, we define below scale-dependent Lyapunov exponents (SDLEs) from the time traces of the CM position \cite{gon23}. 

\begin{center}
\begin{figure}[ht]
\begin{center}
\includegraphics[clip,width=0.97\linewidth]{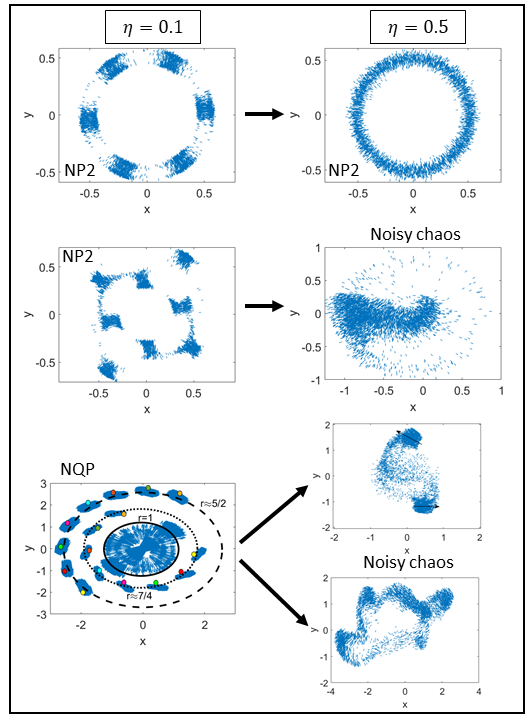}
\\
\end{center}
\caption{{\bf Visual patterns} for $N = 5000, R_0 = 0.472$, $\eta=0.1$ (left column) and $\eta=0.5$ (right column). Left column from top to bottom: $\beta=5\times10^4$ and $\beta = 2\times10^4$ (noisy period-2 cycles), $\beta = 500$ (concentric circular orbits of a noisy quasiperiodic attractor with radii 7/4 and 5/4, points with  different color distinguish times at different iterations). Right column from top to bottom: $\beta = 10^9$ (noisy period-2 cycle), $\beta = 5000$ (noisy chaos: cumulus of particles rotating inside the unit circle), $\beta = 1000$ (two groups of particles oscillate chaotically outside the unit circle while the others remain inside), $\beta = 100$ (deformable group with chaotic dynamics).} \label{fig3}
\end{figure}
\end{center}

As the confinement parameter decreases, different attractors are shown in the left and right columns of Fig.~\ref{fig3} corresponding to two values of the noise strength, $\eta=0.1, 0.5$, respectively ($N= 5000$). For $\eta=0.1$ and $\beta=5\times 10^4$, $\beta=2\times 10^4$, the solutions are period-2 noisy cycles consisting of subgroups of particles forming different patterns alternating densely populated regions with sparsely populated regions. For $\beta=500$, there are noisy  quasiperiodic attractors consisting of a densely populated inner region and a number of orbits comprising different subgroups of particles. This pattern is similar to that in Fig.~\ref{fig1} for $\beta=500$. For the larger noise value $\eta= 0.5$ and $\beta=10^9$, the right column of Fig.~\ref{fig3} shows that the annulus with 6 densely populated regions seen on the left column has been filled. For $\eta= 0.5$ and $\beta=5000$, a rotating chaotic pattern with dense regions appear. These chaotic patterns change their shape for $\beta=1000$ and $\beta=100$.  

\begin{center}
\begin{figure}[ht]
\begin{center}
\includegraphics[clip,width=0.94\linewidth]{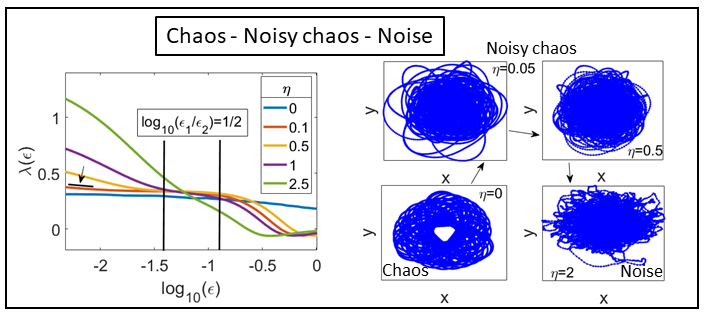}
\\
\end{center}
\caption{SDLE $\lambda(\epsilon)$ vs log$_{10}(\epsilon)$ for different noise values, $\beta=0.01$ and $N=128$. The right panel visualizes the chaotic attractors by showing the trajectories of the CM. Increasing the noise $\eta$ in the sense marked by the arrows transforms a deterministic chaotic attractor into a noisy chaotic attractor until noises predominates at $\eta=2$. \label{fig4}}
\end{figure}
\end{center}

Adding the components of $\mathbf{X}(t)$, we form the time series $x(t)=X(t)+Y(t)$. To calculate the SDLE, we construct the lagged vectors: $\mathbf{X}_\alpha=[x(\alpha),x(\alpha+\tilde{\tau}),..,x(\alpha+(m-1)\tilde{\tau})]$. The simplest choice is $m=2$ and $\tilde{\tau}=1$ (other values can be used, see below). From this dataset, we determine the maximum $\varepsilon_\text{max}$ and the minimum $\varepsilon_\text{min}$ of the distances between two vectors, $\|\mathbf{X}_\alpha-\mathbf{X}_\beta\|$. Our data is confined in $[\varepsilon_\text{min},\varepsilon_\text{max}]$. Let $\varepsilon_0$, $\varepsilon_t$ and $\varepsilon_{t+\Delta t}$ be the average separation between nearby trajectories at times 0, $t$, and $t+\Delta t$, respectively. The SDLE is
 \begin{subequations}\label{eq5}
 \begin{eqnarray}
\lambda(\varepsilon_t)=\frac{\ln\varepsilon_{t+\Delta t}-\ln\varepsilon_t}{\Delta t}. \label{eq5a}
\end{eqnarray}
The smallest possible $\Delta t$ is of course the time step $\tilde{\tau}=1$, but $\Delta t$ may also be chosen as an integer larger than 1. The following scheme yields the SDLE \cite{gao06}: Find all the pairs of vectors in the phase space whose distances are initially within a shell of radius $\epsilon_k$ and width $\Delta\epsilon_k$:
\begin{eqnarray}
\varepsilon_k\leq\|\mathbf{X}_\alpha-\mathbf{X}_\beta\|\leq\varepsilon_k+\Delta\varepsilon_k,\quad k=1,2,\ldots.\label{eq5b}
\end{eqnarray}
We calculate the Lyapunov exponent \eqref{eq5a} as follows:  
\begin{eqnarray}
\lambda(\varepsilon_t)=\frac{\langle\ln\|\mathbf{X}_{\alpha+t+\triangle t}-\mathbf{X}_{\beta+t+\triangle t}\|\!-\!\ln\|\mathbf{X}_{\alpha+t}-\mathbf{X}_{\beta+t}\|\rangle_{k}}{\triangle t}, \label{eq5c}
\end{eqnarray} 
where $\langle\rangle_{k}$ is the average within the shell $(\varepsilon_k,\varepsilon_k+\triangle \varepsilon_k)$. The shell dependent SDLE $\lambda(\varepsilon)$ in Fig.~\ref{fig4} displays the dynamics at different scales for $\tilde{\tau}=9$ and $m=6$ \cite{gao06}. In deterministic chaos, $\lambda(\varepsilon)>0$ presents a plateau with ends $\varepsilon_1< \varepsilon_2\ll 1$, in noisy chaos, this plateau is preceded and succeeded by regions in which $\lambda(\varepsilon)$ decays as $-\gamma\ln\varepsilon$, whereas it shrinks and disappears when noise swamps chaos. As $\eta$ increases, $\lambda(\varepsilon)$ first decays to a plateau for $\eta=0.1$. A criterion to distinguish (deterministic or noisy) chaos from noise is to accept the largest Lyapunov exponent as the positive value at a plateau $(\varepsilon_1,\varepsilon_2)$ satisfying 
\begin{eqnarray}
\log_{10}\frac{\varepsilon_2}{\varepsilon_1}\geq \frac{1}{2}.  \label{eq5d}
\end{eqnarray}\end{subequations}
For $\eta=0.5$, the region where $\log_{10}(\varepsilon_2/\varepsilon_1)= 1/2$ is marked in Fig.~\ref{fig4} by vertical lines. Plateaus with smaller values of $\log_{10}(\varepsilon_2/ \varepsilon_1)$ or their absence indicate noisy dynamics \cite{gao06}. This occurs for $\eta=1$. 

\begin{center}
\begin{figure}[ht]
\begin{center}
\includegraphics[clip,width=0.97\linewidth]{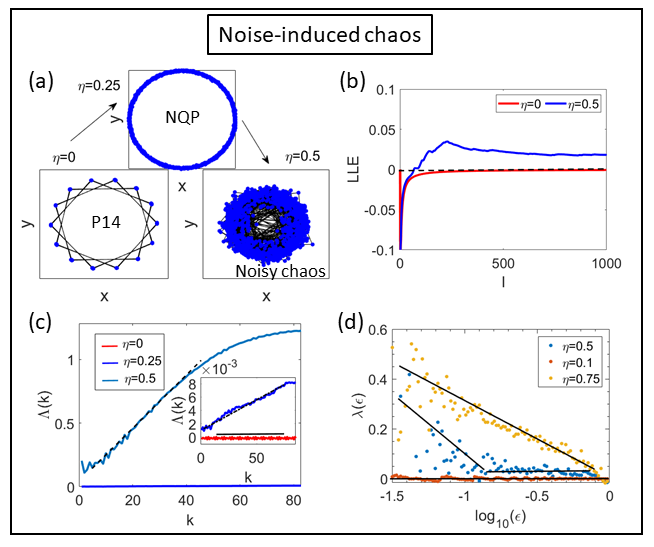}
\\
\end{center}
\caption{Noise can induce chaos starting from a period-14 cycle in the deterministic case:  {\bf (a)} for $\eta=0.25$ the periodic solution has become a noisy annular quasiperiodic solution, which turns into noisy chaos for $\eta=0.5$ (left panel). Here $\beta=99.52$, $N=128$.  {\bf (b)} LLEs $\lambda_1=0$ for $\eta=0$ and $\lambda_1=0.025$ for $\eta=0.5$ calculated by the BA \cite{ben80}. {\bf (c)} $\Lambda(k)$ vs $k$ as calculated by the GZA \cite{gao94} of Eq.~\eqref{eq6}, which yields $\lambda_1=0.023$ for $\eta=0.5$.  {\bf (d)}  The SDLE (with $m=6$, $\tilde{\tau}=9$) distinguishes deterministic periodic solutions (zero slope) from NQP ($\eta=0.1$) and noisy chaos ($\eta=0.5$) while yielding the same LLE as the GZA. \label{fig5}}
\end{figure}
\end{center}

Fig.~\ref{fig5} shows that noise can induce chaos. Starting from a period-14 cycle in the deterministic case, the periodic solution becomes a noisy annular quasiperiodic solution for $\eta=0.1$, which then turns into noisy chaos for $\eta=0.5$; see Fig.~\ref{fig5}(a).  We have calculated the LLEs by using the BA, Fig.~\ref{fig5}(b), Gao-Zheng algorithm (GZA) \cite{gao94}, Fig.~\ref{fig5}(c), and by the SDLE, Fig.~\ref{fig5}(d). The GZA consists of constructing the quantity $\Lambda(k)$, 
\begin{eqnarray}
\Lambda(k)= \left\langle\ln\frac{\|\mathbf{X}_{\alpha+k}-\mathbf{X}_{\beta+k}\|}{\|\mathbf{X}_\alpha-\mathbf{X}_\beta\|}\right\rangle,  \label{eq6}
\end{eqnarray}
whose slope near the origin gives the LLE \cite{gao94}, as shown in the central panel of Figure \ref{fig5}. In Eq.~\eqref{eq6}, the brackets indicate ensemble average over all vector pairs with $\|\mathbf{X}_\alpha-\mathbf{X}_\beta\|<r^*$ for an appropriately selected small distance $r^*$. The LLEs are $\lambda_1=0$ (P-14 and NQP) and $\lambda_1=0.023$ for $\eta=0.5$.

For the SDLE to produce accurate values of the LLE, we need sufficient lagged coordinates for safely reconstructing the chaotic attractor. This is achieved if the dimension of the lagged vectors is twice the fractal dimension $D_0$ or larger \cite{ott93}. Thus, we have used $m=6$ lagged coordinates and $\tilde{\tau}=9$ in Fig.~\ref{fig5}(c) and then GZA and SDLE yield the same values of the LLE. Moreover, P-14 and NQP solutions have zero and negative slope of $\lambda(\epsilon)$ at the beginning of the $\epsilon$ interval, denoting deterministic and noisy solutions, respectively.

\section{Scale-free chaos phase transition}\label{sec4}
\begin{center}
\begin{figure}[ht]
\begin{center}
\includegraphics[clip,width=8.5cm]{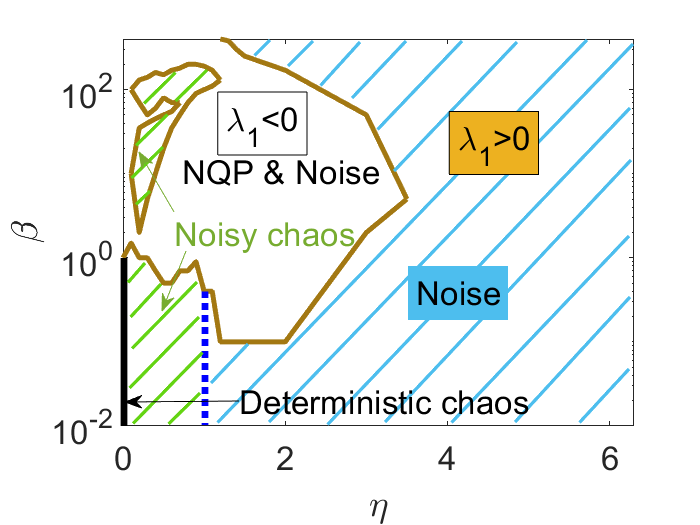}
\\
\end{center}
\caption{{\bf Phase diagram $\beta$ vs $\eta$ for $N=500$.}}
 \label{fig6}
\end{figure}
\end{center}

In the previous sections, the numerical simulations of the 2D HCVM have shown the existence of different attractors for fixed values of the number of particles $N$, confinement strength $\beta$ and noise $\eta$. Fig.~\ref{fig6} shows the phase diagram of different attractors on the plane $(\beta,\eta$). As in the case of the 3D HCVM \cite{gon23,gon23mf,gon23arxiv}, as $\beta\to 0$ and $N\to\infty$, there is a phase transition of scale-free chaos. At finite $N$, scale free means that the correlation length is proportional to the size of the swarm, so that all other characteristic lengths are not important. There are several critical curves on the phase diagram that tend to $\beta=0$ at the same rate as $N\to\infty$. We describe here the critical curve $\beta_c(N;\eta)$ separating single cluster from multicluster chaotic attractors \cite{gon23}. There are other critical curves that have been studied for the 3D HCVM at fixed $\eta$ and increasing valued of $N$: (i) The critical curve $\beta_m(N;\eta)$ of local maxima of the susceptibility as a function of $\beta$; (ii) the curve $\beta_i(N;\eta)$ of inflection points of the susceptibility vs $\beta$; (iii) the curve $\beta_0(N;\eta)$ separating regions of zero and positive LLE. These curves satisfy the relation $\beta_0(N;\eta)<\beta_c(N;\eta)<\beta_i(N;\eta)<\beta_m(N;\eta)$ \cite{gon23,gon23arxiv}. In 3D, all these critical curves converge to $\beta=0$ at the same rate as $N\to\infty$ \cite{gon23,gon23arxiv}. In 2D, we have not found a curve $\beta_0(N;\eta)$ because the algorithms used to compute the LLE do not converge at values of $\beta$ that are too small: for all $\beta<\beta_c(N;\eta)$ within the range of convergence of the BA, the LLE is positive.

\subsection{Correlation functions and power laws}
Let us first define the static and dynamic correlation functions, the correlation length and time and the susceptibility with the corresponding power laws and critical exponents. The dynamic connected correlation function (DCCF) is  \cite{att14,cav18}
\begin{subequations}\label{eq7}
\begin{eqnarray}\
C(r,t)=\!\left\langle\! \frac{\sum_{i=1}^{N}\!\sum_{j=1}^{N}\delta\hat{\mathbf{v}}_i(t_0)\!\cdot\!\delta\hat{\mathbf{v}}_j(t_0\!+t)\,\delta[r-r_{ij}(t_0,t)]}{\sum_{i=1}^{N}\sum_{j=1}^{N}\delta[r-r_{ij}(t_0,t)]}\! \right\rangle_{t_0}\!,\quad C(r)=C(r,0), \label{eq7a}\end{eqnarray}
where
\begin{eqnarray}
&&\delta\hat{\mathbf{v}}_i\!=\frac{\delta\mathbf{v}_i}{\sqrt{\frac{1}{N}\sum_k \delta\mathbf{v}_k\cdot\delta\mathbf{v}_k}},\quad \delta\mathbf{v}_i=\mathbf{v}_i - \mathbf{V},\label{eq7b}\\
&&r_{ij}(t_0,t)=|\mathbf{r}_i(t_0)-\mathbf{r}_j(t_0+t)|,\,  \mathbf{r}_i(t_0)=\mathbf{x}_i(t_0) - \frac{1}{N}\sum_{j=1}^N\mathbf{x}_j(t_0),   \label{eq7c}\\
&&\langle f\rangle_{t_0}=\frac{1}{t_{max}-t}\sum_{t_0=1}^{t_{max}-t}f(t_0,t).\label{eq7d} 
\end{eqnarray}\end{subequations}
In Eq.~\eqref{eq7a}, $\delta(r-r_{ij})=1$ if $r<r_{ij}<r+dr$ and zero otherwise, and $dr$ is the space binning factor. The averages are over time and over five independent realizations corresponding to five different random initial conditions during 10000 iterations \cite{gon23}. The SCCF is the equal time connected correlation function $C(r)=C(r,0)$ given by Eq.~\eqref{eq7a}. Note that $C(\infty)\propto |\sum_{i=1}^N\delta\mathbf{\hat{v}}_i|^2=0$. The correlation length $\xi$ can be defined as the first zero of $C(r)$, $r_0$, corresponding to the first maximum of the cumulative correlation function \cite{att14}:
\begin{eqnarray}
&&Q(r)=\!\left\langle \frac{1}{N}\sum_{i=1}^{N}\sum_{j=1}^{N} \delta\hat{\mathbf{v}}_i\!\cdot\!\delta\hat{\mathbf{v}}_j\theta(r-r_{ij}(t_0,0))\right\rangle_{t_0}\!, \quad\chi= Q(\xi),\label{eq8}\\ 
&&\xi=\mbox{argmax}_rQ(r),\, C(\xi)=0\,\mbox{ with }\, C(r)>0,\, r\in(0,\xi), \quad\nonumber
\end{eqnarray}
where $\theta(x)$ is the Heaviside unit step function. For $r$ larger than the swarm size, $Q(r)=\langle |\sum_{i=1}^N \delta\mathbf{\hat{v}}_i|^2 \rangle_{t_0}/N=0$. The susceptibility $\chi$ is the value of $Q(r)$ at its first maximum, as in Ref.~\onlinecite{att14}. Alternatively, we can use the Fourier transform of Eq.~\eqref{eq7a},
\begin{eqnarray}
\hat{C}(k,t)\!=\!\left\langle\! \frac{1}{N}\!\sum_{i,j =1}^{N}\!\!\frac{\sin(kr_{ij}(t_0,t))}{kr_{ij}(t_0,t)}\delta\hat{\mathbf{v}}_i(t_0)\!\cdot\!\delta\hat{\mathbf{v}}_j(t_0+t)\! \!\right\rangle_{t_0}\quad\label{eq9}
\end{eqnarray}
and define the critical wavenumber $k_c=$argmax$_k\hat{C}(k,0)$, the susceptibility as $\chi=\max_k\hat{C}(k,0)$, and the correlation length as $\xi=1/k_c$ \cite{cav17,cav18,gon23}. It turns out that $k_c\propto 1/r_0$ on critical curves and we can use either the real-space or the Fourier space SCCF to find correlation length and susceptibility. On the critical curves where the correlation length is proportional to the swarm size, the correlation length and the  susceptibility obey power laws with critical exponents $\nu$ and $\gamma$, respectively:
\begin{subequations}\label{eq10}
\begin{eqnarray}
\xi\sim \beta^{-\nu},\quad \chi\sim\beta^{-\gamma}. \label{eq10a}
\end{eqnarray}
As $N\to\infty$, the confinement on the critical curve $\beta$ tends to zero and correlation length and susceptibility diverge \cite{hua87,ami05}. Similarly, the polarization, i.e., the average speed of the CM velocity, tends to zero as a power law with critical exponent $b$:
\begin{eqnarray}
\langle W\rangle_t\sim \beta^b,\quad W=\left|\frac{1}{N}\sum_{j=1}^N\frac{\mathbf{v}_j}{|\mathbf{v}_j|}\right|=\frac{\left|\sum_{j=1}^N \mathbf{v}_j\right|}{Nv_0}.\label{eq10b}
\end{eqnarray}\end{subequations}

For the DCCF, the dynamic scaling hypothesis implies 
\begin{eqnarray}
&&\frac{\hat{C}(k,t)}{\hat{C}(k,0)}= f\!\left(\frac{t}{\tau_k},k\xi\right)\!= g(k^zt,k\xi); \nonumber\\
&& g(t)=\frac{\hat{C}(k_c,t)}{\hat{C}(k_c,0)};\quad \tau_k=k^{-z}\phi(k\xi). \label{eq11}
\end{eqnarray}
Here $z$ is the dynamic critical exponent and the correlation time $\tau_k=k^{-z}\phi(k\xi)$ of the normalized DCCF (NDCCF) \eqref{eq11} at wavenumber $k$ obtained by solving the equation: \cite{cav17,gon23}
\begin{eqnarray}
\sum_{t=0}^{t_{max}} \frac{1}{t}\,\sin\!\left(\frac{t}{\tau_k}\right) f\!\left(\frac{t}{\tau_k},k\xi\right)\! = \frac{\pi}{4}.  \label{eq12}
\end{eqnarray}
At $k_c=$argmax$_k\hat{C}(k,0)$, the correlation time $\tau_{k_c}$ is a function of $\beta$, $\eta$ and $N$. For fixed $N$ and $\eta$, there is a value of $\beta=\beta_c$ at which $\tau_{k_c}$ reaches a minimum. This minimum correlation time corresponds to the smallest time $t_m(\beta;\eta,N)$ at which $\hat{C}(k_c,t)=0$ \cite{gon23}. See Figure 5(a) of Ref.~\cite{gon23} for the 3D HCVM. 

\begin{center}
\begin{figure}[ht]
\begin{center}
\includegraphics[clip,width=\linewidth]{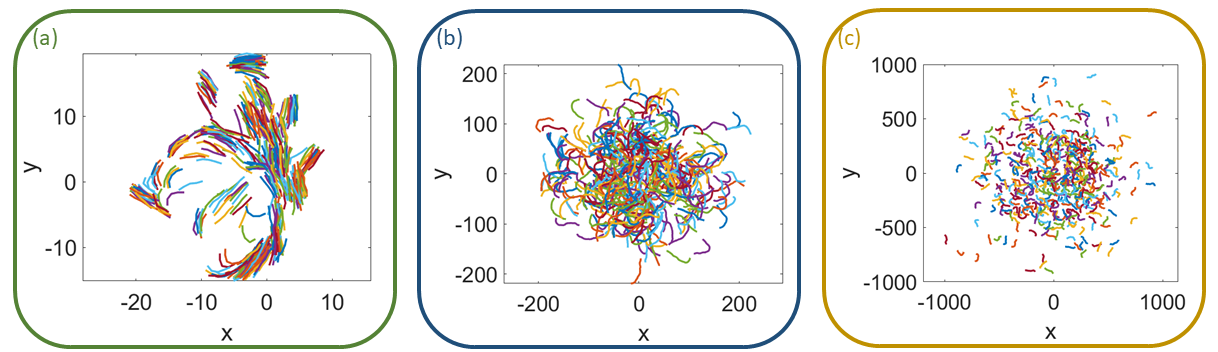}
\\
\end{center}
\caption{{\bf Visual patterns of the chaotic attractors in (a)} chaotic multicluster phase for $\beta=0.1$, {\bf (b)} scale-free chaos phase transition for $\beta_c=0.0003$ and {\bf (c)} chaotic single cluster phase for $\beta=10^{-5}$. Here $\eta=0.5$ and $N=500$.}
 \label{fig7}
\end{figure}
\end{center}

\subsection{Scale free chaos and critical exponents}
For finite $N$, the 2D HCVM exhibits a region of chaos and noisy chaos for small values of $\beta$ and $\eta$ as shown in Fig.~\ref{fig6}. Deep into this region, there is a transition between two types of chaotic attractors: multicluster chaos, in which the swarm is split into several groups of particles as in Fig.~\ref{fig7}(a), and single cluster chaos (only one group) as in Fig.~\ref{fig7}(b) (at the transition value $\beta = \beta_c$), and in Fig.~\ref{fig7}(c) ($\beta<\beta_c$). 

\begin{center}
\begin{figure}[ht]
\begin{center}
\includegraphics[clip,width=\linewidth]{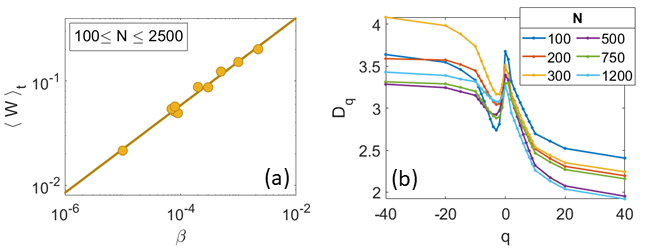}
\\
\end{center}
\caption{{\bf Scale-free chaos phase transition for $\beta\rightarrow0$, $N\rightarrow\infty$ and $\eta=0.5$.} {\bf (a)} Polarization order parameter vs $\beta$ power law yielding the critical exponent $b=0.42\pm0.02$. {\bf (b)} Multifractal dimension $D_q$ calculated from the chaotic attractor center of mass.}
 \label{fig8}
\end{figure}
\end{center}

The polarization order parameter goes to zero as $\beta\to 0$ following the power law Eq.~\eqref{eq10b} with $b=0.42\pm 0.02$; see Fig.~\ref{fig8}(a).  Moreover, the chaotic attractor is multifractal \cite{ott93}, as shown by Fig.~\ref{fig8}(b). Then some regions of the chaotic attractor are visited more often than others, which indicates that different length and timescales coexist within the attractor. This is made manifest by calculating the multifractal dimension $D_q$. After a long transient (30 000 time steps), a set of $M$ values of the CM position $\vec{x}_i = X(t_i)+Y(t_i)$, $i = 1, \ldots,M$, form a Poincar\'e map of the attractor. The multifractal dimension is
\begin{eqnarray}
D_q = \frac{1}{q-1} \lim_{r \rightarrow 0} \frac{\ln [C_q(r)]}{\ln (r)},\quad
C_q(r)=  \frac{1}{M} \sum_{i=1}^M  \left[ \frac{1}{M} \sum_{j=1}^M \theta(r-|\vec{x}_i-\vec{x}_j|) \right]^{q-1},  \label{eq13}
\end{eqnarray}
where $\theta(x)$ is the Heaviside unit step function, $M\approx 70000$, and $C_q(r)$ is the generalized correlation function. $D_0$, $D_1$ and $D_2$ are the box counting (capacity) dimension, the information dimension and the correlation dimension, respectively. As we vary $q$, different regions of the attractor will determine $D_q$. $D_\infty$  corresponds to the region where the points are mostly concentrated, while $D_{-\infty}$ is determined by the region where the points have the least probability to be found. If $D_q$ is a constant for all $q$, the CM trajectory will visit different parts of the attractor with the same probability and the point density is uniform in the Poincar\'e map. This type of attractor is called trivial, whereas a non constant $D_q$ characterizes a nontrivial attractor with multifractal structure. Fig.~\ref{fig8}(b) shows that the box-counting dimension $D_0$ and $D_q$ for $q>0$ undergo a downward trend with increasing $N$ (decreasing $\beta_c$). Then the dimension of the more commonly visited sites of the attractor decreases. The chaotic attractor remains multifractal when $\beta\to 0$ as $N\to\infty$ and chaos disappears: different time scales persist \cite{cen10}. 

\begin{center}
\begin{figure}[ht]
\begin{center}
\includegraphics[clip,width=0.9\linewidth]{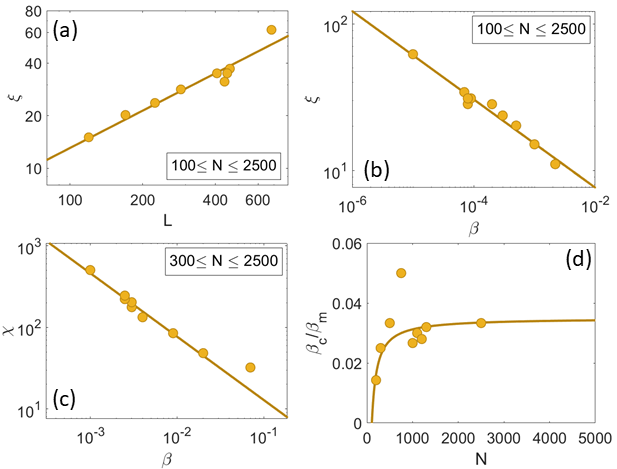}\\
\end{center}
\caption{{\bf Scale-free chaos phase transition for $\beta\rightarrow0$, $N\to\infty$ and $\eta=0.5$. (a)} Scale free $\xi\sim L$ ($\xi=r_0$). {\bf (b)} Correlation length vs $\beta$ power law producing the critical exponent $\nu=0.30\pm 0.02$. {\bf (c)} Susceptibility vs $\beta$ power law from the critical curve $\beta_m(N;\eta)$ yielding $\gamma=0.78\pm 0.05$. {\bf (d)} Ratio of $\beta_c/\beta_m$ vs $N$ giving $\beta_c/\beta_m\to 0.035$ as $N\to\infty$. \label{fig9}}
\end{figure}
\end{center}

As explained above, $\beta_c=$argmin$_\beta\tau_{k_c}(\beta;\eta,N)$, where $\tau_{k_c}$ is the minimum correlation time given by the solution of Eq.~\eqref{eq12} for fixed $\eta$ and $N$. At $\beta_c$, the correlation length and the size of the swarm are proportional for different values of $N$ as shown by Fig.~\ref{fig9}(a). This indicates that the chaotic attractor of the HCVM is scale free for the curve $\beta_c(\eta;N)$.  

\begin{center}
\begin{figure}[ht]
\begin{center}
\includegraphics[clip,width=0.9\linewidth]{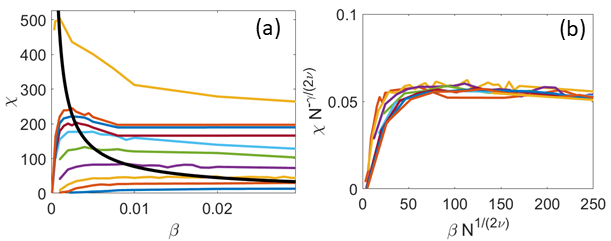}\\
\end{center}
\caption{{\bf Susceptibility. (a)} Power law of the susceptibility vs $\beta$. {\bf (b)} Scaled susceptibility vs scaled $\beta$ yielding $\gamma=0.78\pm 0.05$ as $N\to\infty$. The colors correspond to different values of $N$ in the interval $300\leq N\leq 2500$. \label{fig10}}
\end{figure}
\end{center}

Figs.~\ref{fig9}(b) and \ref{fig9}(c) show that, for increasing $N\to\infty$, correlation length and susceptibility scale with $\beta=\beta_c$ (with $\lim_{N\to\infty}\beta_c(\eta;N)= 0$) following the power laws \eqref{eq10a} with critical exponents $\nu=0.30\pm 0.02$ and $\gamma=0.78\pm0.05$, respectively. While the correlation length and time are calculated using numerical simulations of the HCVM on the critical curve $\beta_c(\eta;N)$, the susceptibility does not tend to a definite value as $N$ increases due to the very small values of $\beta_c$ for large $N$ (which are much smaller for the 2D HCVM than the corresponding values for the 3D HCVM). To fix this problem of the 2D HCVM, we recall that the scale-free-chaos phase transition in the 3D HCVM has an extended critical region which collapses to $\beta=0$ at the same rate for all values of the noise as $N\to\infty$ \cite{gon23}. The critical curve with larger values of $\beta$ is $\beta_m(\eta;N)$, corresponding to the local maximum of the susceptibility $\chi$ with respect to $\beta$ (at fixed $N$ and noise) calculated as in $\chi$ in Equations~\eqref{eq8}, $\chi=Q(r_0)$, or as $\chi=\max_k\hat{C}(k,0)$ \cite{gon23}. The curve $\beta_m(\eta;N)$ is also scale free and it yields scaling laws as in Equations~\eqref{eq10}. Figs.~\ref{fig10}(a) and \ref{fig10}(b) illustrate that the power law Eq.~\eqref{eq10a} for the susceptibility $\chi=Q(r_0)$ at $\beta=\beta_m$. Moreover, Fig.~\ref{fig9}(d) shows that the curves $\beta_c$ and $\beta_m$ tend toward $\beta=0$ at the same rate: $\beta_c/\beta_m\to 0.035$ as $N\to\infty$. Thus, we calculate the critical exponent of the susceptibility using $\beta_m(\eta;N)$ instead of $\beta_c(\eta;N)$, thereby obtaining Fig.~\ref{fig9}(c). 

\begin{center}
\begin{figure}[ht]
\begin{center}
\includegraphics[clip,width=0.87\linewidth]{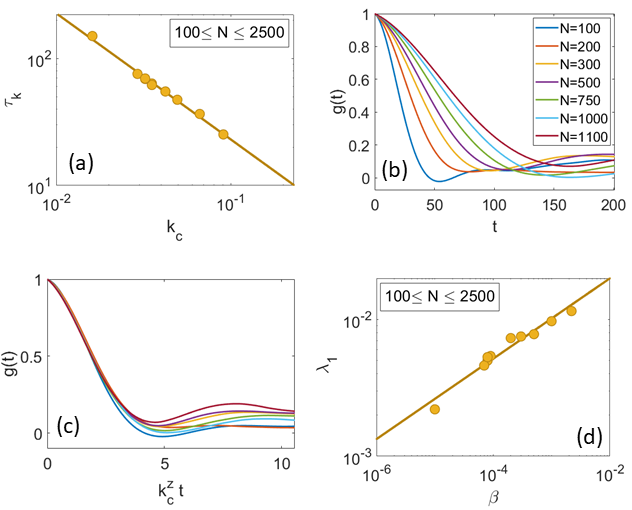}\\
\end{center}
\caption{{\bf Scale-free chaos phase transition for $\beta\rightarrow0$, $N\to\infty$ and $\eta=0.5$. (a)} Correlation time vs $k_c=1/\xi$ for $100\leq N\leq2500$ yielding $z_\text{LS}=z_\text{RMA}= 0.99\pm0.03$. {\bf (b)} Normalized dynamic connected correlation function (NDCCF) $g(t)=\hat{C}(k_c,t)/\hat{C}(k_c,0)$ for $\beta=\beta_c(N;\eta)$. {\bf (c)} Visual collapse of the NDCCF as a function of $k_c^zt$ for $z\approx1$. {\bf (d)} LLE vs $\beta$ with critical exponent $\varphi=0.29\pm0.02= z\nu$. \label{fig11}}
\end{figure}
\end{center}

The dynamical critical exponent relating correlation time and length, $\tau_k\sim\xi^z\sim k_c^{-z}$, is $z=0.99\pm 0.03$; see Fig.~\ref{fig11}(a). For different values of $N$, the NDCCF $g(t)$ of Eq.~\eqref{eq11} is shown in Fig.~\ref{fig11}(b). As shown in Fig.~\ref{fig11}(c), the different NDCCF curves collapse into a single curve for small values of the scaled time $k_c^zt$. The same collapse of the NDCCF as a function of $k_c^zt$ only for $0<k_c^zt<4$ occurs using data from natural swarms, as shown in Figures 2a and 2b of Ref.~\onlinecite{cav17} for $z=1.2$. We ascribe the partial collapse of the NDCCF to the multifractality of the chaotic attractors shown in Fig.~\ref{fig8}(b), which indicates that different length and time scales persist for all the critical values of $\beta$. Thus, a single rescaling of time as in Fig.~\ref{fig11}(c) cannot collapse the full NDCCF in our simulations. Furthermore, Fig.~\ref{fig11}(d) indicates that the LLE scales with $\beta=\beta_c$ as 
\begin{eqnarray}
\lambda_1\sim \beta^\varphi,    \label{eq14}
\end{eqnarray}
with $\varphi=0.29\pm 0.02=z\nu$. As for the 3D HCVM, these scaling laws illustrate the scale-free-chaos phase transition \cite{gon23}.  

\begin{center}
\begin{figure}[ht]
\begin{center}
\includegraphics[clip,width=0.87\linewidth]{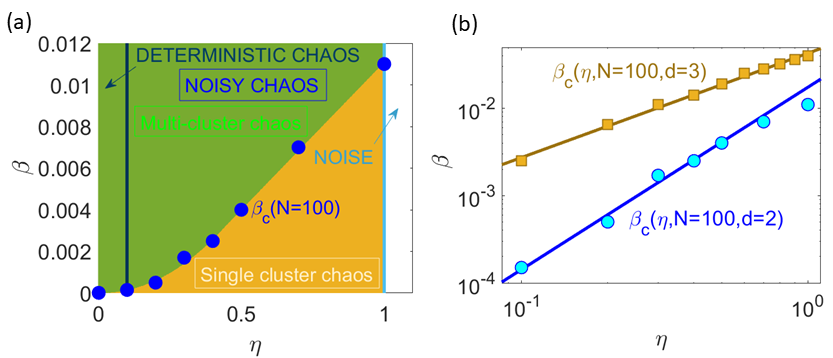}\\
\end{center}
\caption{{\bf Scale-free chaos phase transition for $\beta\rightarrow0$ and $\eta\rightarrow0$.} {\bf (a)} Critical curve $\beta_c(\eta,N)$ on the parameter space $(\beta,\eta)$ separating phases of single cluster from multicluster chaos for $N=100$ and $0.1\leq\eta\leq 1.0$. The power law fitting this curve is  $\beta_c=C_cN^{-\frac{1}{2}}\eta^{m_c}$ with $C_c/\sqrt{100}=0.017\pm0.003$ and $m_c=2.1\pm0.1$ for $0.1\leq\eta\leq0.5$ ($\eta=0.7$ and $\eta=1$ are outside the fit). {\bf (b)} Comparison of the curve $\beta_c$ vs $\eta$ for $N=100$ for the 2D HCVM (blue curve with slope $m_c=2.1$) and 3D HCVM (brown curve with slope $m_c=1.2$). \label{fig12}}
\end{figure}
\end{center}

\section{Unfolding of the phase transition at small noise}\label{sec5}
\begin{center}
\begin{figure}[ht]
\begin{center}
\includegraphics[clip,width=0.9\linewidth]{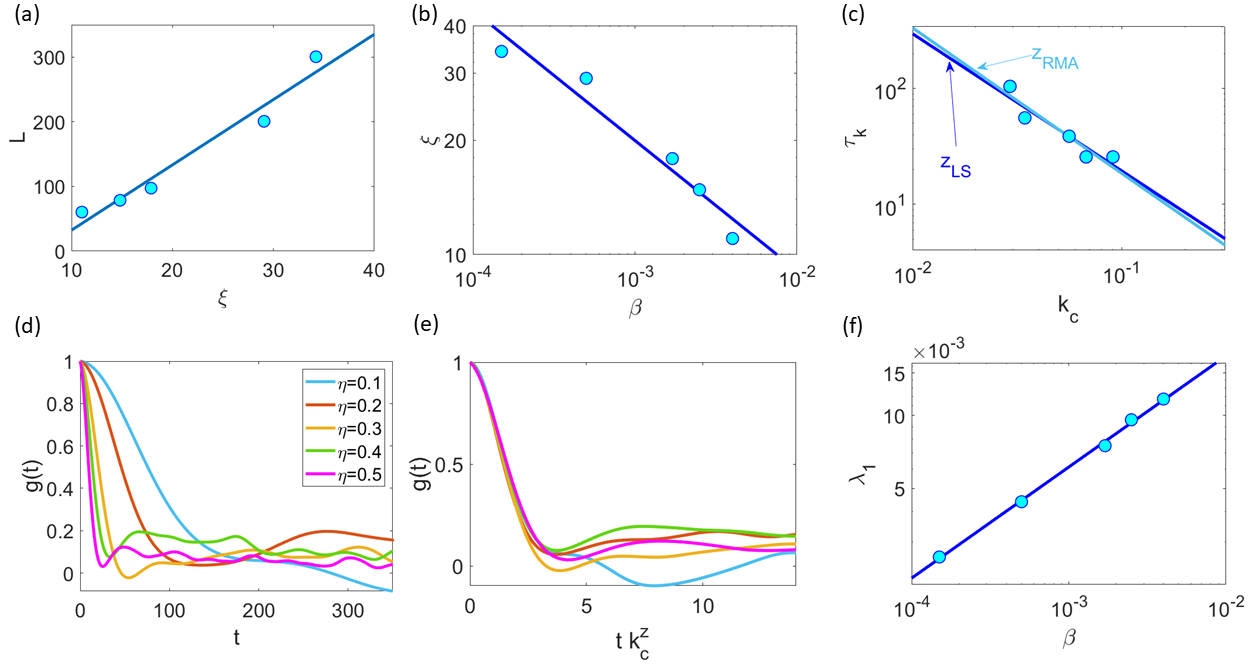}\\
\end{center}
\caption{{\bf Scale-free chaos phase transition for $\beta\rightarrow0$ and $\eta\rightarrow0$, for $N=100$ and $0.1\leq\eta\leq0.5$.} {\bf (a)} Scale free property $\xi\sim L$. {\bf (b)} Power law of $\xi$ vs $\beta$ with $\nu=0.34\pm0.05$. {\bf (c)} Dynamical critical exponent from correlation time vs $k_c=1/\xi$: $z_\text{LS}=1.18\pm0.11$ and $z_\text{RMA}= 1.24\pm0.12$. {\bf (d)} Normalized dynamic connected correlation function (NDCCF) $g(t)=\hat{C}(k_c,t)/\hat{C}(k_c,0)$ for $\beta=\beta_c(N;\eta)$. {\bf (e)} Visual collapse of the NDCCF as a function of $k_c^zt$ for $z\approx1.15$. {\bf (f)} LLE vs $\beta$ yielding $\varphi=0.46\pm0.01\approx z\nu$. \label{fig13}}
\end{figure}
\end{center}

The scale-free-chaos phase transition in the 3D HCVM is such that all critical curves tend to $\beta=\eta=0$ for finite number of particles \cite{gon23arxiv}. There are power laws in $\eta$ that allow the calculation of critical exponents without having to increase $N$ as we did in Section \ref{sec4}. The point $\beta=\eta=0$ acts as an organizing center of codimension two for the phase transition in a sense that reminds of singularity theory \cite{gol85}. The critical curves including $\eta=0$ and $\beta_c(\eta;N)$ issue from the organizing center at finite $N$ and the attractors in the regions between these lines have specific properties (non-chaotic, single-cluster chaos, multicluster chaos, deterministic chaos, etc). Two parameters, $\beta$ and $\eta$, unfold these behaviors as they take on nonzero values. As $N\to\infty$, all critical curves tend to $\beta=0$ (scale-free-chaos phase transition). On the critical curve $\beta_c$, the 2D power law is \cite{gon23arxiv}
\begin{subequations} \label{eq15}
\begin{eqnarray}
\beta_c(N;\eta)= C_c N^{-\frac{1}{2\nu}}\eta^{m _c}.   \label{eq15a}
\end{eqnarray}
For $N=100$, $C_c N^{-\frac{1}{2\nu}}=0.017\pm 0.003$ and $m_c=2.1\pm 0.1$. Taking into account the power law \eqref{eq10a} for the correlation length, the previous equation can be rewritten as
\begin{eqnarray}
\xi\sim N^\frac{1}{2}\eta^{-\nu m_c}. \label{eq15b}
\end{eqnarray}\end{subequations}
This expression can be used to obtain $\nu$ for fixed $N$. Consider the critical curve $\beta_c(\eta;N)$ for $N=100$ in Fig.~\ref{fig12}(a) and the points on it corresponding to noise values $0.1\leq\eta\leq 0.5$. Fig.~\ref{fig12}(b) compares the curves $\beta_c(\eta)$ at $N=100$ for the 2D and 3D HCVM. We observe that the 2D values of critical confinement become much smaller than the corresponding 3D values as the noise decreases. Fig.~\ref{fig13}(a) shows that the critical curve is indeed scale free as the size $L$ of the swarm is proportional to the correlation length $\xi$. Fig.~\ref{fig13}(b) plots $\xi$ as a function of $\beta$ on the critical curve $\beta_c$ and yields $\nu=0.34\pm 0.05$. This value is compatible with $\nu=0.30\pm 0.02$ obtained in Section \ref{sec4} by calculating several points on the curve $\beta_c(\eta;N)$ for fixed $\eta=0.5$ and $100\leq N\leq 2500$.

We can also obtain the dynamical critical exponent by scaling time in the graph of the NDCCF $g(t)$ to $k_c^zt$, with $k_c=1/\xi$, as shown in  Fig.~\ref{fig13}(c). Least square (LS) and reduced major axis (RMA) regressions \cite{cav23} shown in this figure produce similar values of the dynamical critical exponent, $z_\text{LS}=1.18\pm0.11$ and $z_\text{RMA}=1.24\pm0.12$, respectively. Figs.~\ref{fig13}(d) and \ref{fig13}(e) shows that the curves for different values of $\eta\in[0.1,0.5]$ visually collapse when time is rescaled with $z\approx 1.15$. 

As the number of particles goes to infinity, the LLE tends to zero according to the scaling law Eq.~\eqref{eq14}. Using noise values between 0.1 and 0.5 for $N=100$, we find $\varphi=0.46\pm 0.01$; see Fig.~\ref{fig13}(f). This exponent is compatible with the dynamic relation $\varphi=z\nu$ found for the 3D HCVM in Ref.~\cite{gon23}. However, this value is different from that obtained in Section \ref{sec4} by considering increasing values of $N$ at fixed $\eta = 0.5$.

\section{Discussion}\label{sec6}
We have numerically simulated the 2D HCVM for different values of the noise strength $\eta$ and the number of particles $N$. As the confinement parameter decreases, the model exhibits periodic attractors with periods 2, 4, and so on, followed by quasiperiodic attractors and, eventually, chaotic attractors with different shapes that occupy regions of finite area and comprise one or several groupings. The noise modifies these attractors. Using SDLEs, we can distinguish essentially deterministic chaos from chaos modified by noise (noisy chaos) and from predominantly noisy signals \cite{gao06}; see Fig.~\ref{fig4} and the phase diagram in Fig.~\ref{fig6}. Within the parameter region of deterministic and noisy chaotic solutions, there is a phase transition at $\beta=0$, $N=\infty$. To study this transition using numerical simulations, we use the finite size scaling and dynamical scaling hypotheses \cite{hua87,ami05,hoh77,att14,cav17,cav18}: on critical curves, the characteristic timescale, and the static and dynamic connected correlation functions depend on the control parameters $\beta$ and $\eta$ only through the correlation length $\xi$. On critical curves, the swarm size and the correlation length are proportional, hence $\xi\propto N^\frac{1}{d}$ ($d$ is the space dimension, 2 in the present work). Finite-size scaling allows us to extrapolate power laws of macroscopic variables obtained for finite $N$ to the case of infinitely many particles, which characterize phase transitions \cite{ami05}. 

For finite $N$, there is an extended critical region comprising different critical curves on the plane $(\eta,\beta)$ that converge to $\beta=0$ at the same rate as $N\to\infty$. On the critical curves, the chaotic attractors are multifractal, spanning many different length scales, even as $N$ increases. Compared to the 3D case \cite{gon23arxiv}, the confinement parameter on the critical curves is much smaller. In fact, the BA algorithm used to find the LLE ceases to converge before we succeed in finding a zero value of the LLE. This precludes studying the curve $\beta_0(N;\eta)$ separating chaotic from non-chaotic regions, which necessarily lies below the critical curve $\beta_c(N;\eta)$ separating single from multicluster chaos. For values on this critical curve, the chaotic attractor occupies a connected finite region of space. It is shaped as a condensed core plus particles that enter and exit the core, as shown in Fig.~\ref{fig7}(b) and \ref{fig7}(c). This is similar to observations of midge swarms in the laboratory \cite{sin17} 

With these limitations, we have used our numerical simulations to evaluate static and dynamical critical exponents of the scale-free-chaos phase transition in 2D. We have characterized the critical curve $\beta_c(N;\eta)$ and computed the static critical exponents $\nu$ and $\gamma$ associated to correlation length and susceptibility, respectively, as well as the dynamical critical exponents $z$ and $\varphi$. To derive the power laws and calculate critical exponents, we have used two limiting procedures. In the traditional one, we keep $\eta$ fixed and calculate $\beta_c(N;\eta)$ for increasing values of $N$, using to calculate $\gamma$ (corresponding to the susceptibility) that the curve of local maxima of the susceptibility, $\beta_m(N;\eta)$, collapses to $\beta=0$ at the same rate as $\beta_c(N;\eta)$; see Fig.~\ref{fig9}(d). The other procedure follows from the fact that all critical curves tend to $\eta=0$, $\beta=0$, which produces power laws in $\eta$ for fixed $N$ \cite{gon23arxiv}. Using this second procedure as in Eq.~\eqref{eq15b}, we have found a value of $\nu$ (corresponding to the correlation length) compatible with that obtained by the first procedure.  For the static critical exponent $\gamma$, the proportionality coefficient between $\chi N^{-\gamma/(d\nu)}$ and $\eta$ is independent of $\eta$ and therefore we cannot use the power laws in $\eta$ to find it \cite{gon23arxiv}.

The exponent $z$ characterizes the collapse of the normalized dynamical correlation function when expressed as a function of $k_c^zt=t/\xi^z$. Unlike the case of critical dynamics near equilibrium, the NDCCF collapses only for an interval of small rescaled times (of width about 4), not for all rescaled times. This partial collapse is also observed in experiments \cite{cav17}, but not in models that try to explain experiments by an ordering phase transition between spatially homogeneous phases \cite{cav17,cav23}. We ascribe this partial collapse of the NDCCF to the multifractal nature of chaotic attractors, which contain many different length and time scales \cite{gon23}. As the scale-free-chaos phase transition occurs as $\beta\to 0$, $N\to\infty$, the LLE of the chaotic attractor tends to zero and it follows a new power law while doing so. This power law has a critical exponent $\varphi$, which we have also calculated. 

How can we compare our results with observations of swarms of midges? On the qualitative side, the shape of the swarm is similar to that found in our numerical simulations: a condensed core with a gas of particles (insects) going in and out of the core \cite{sin17}. Furthermore, the partial collapse of the NDCCF for a finite interval of scaled time having about 4 units  of width agrees with observations \cite{cav17}, whereas theories based on the ordering transition in different models produce a collapse of the NDCCF for all values of the scaled time \cite{cav17,cav23}. On the quantitative side, the numerical simulations of the 2D HCVM produce static exponents $\nu=0.30\pm 0.02$, $\gamma=0.78\pm 0.05$ and dynamical exponents $z=0.99\pm 0.03$, $\varphi=0.29\pm 0.02$ (Section \ref{sec4}). Numerical simulations of the 3D HCVM (using the same critical curve $\beta_c(N;\eta)$ as in the present work) yield $\nu=0.436\pm 0.009$, $\gamma=0.92\pm 0.05$, $z= 1.01 \pm 0.01$, and $\varphi=z\nu$ \cite{gon23}. Critical exponents measured from natural swarms are $\nu= 0.35\pm 0.10$, $\gamma=0.9\pm 0.2$ (Ref.~\onlinecite{att14}), and $z=1.37\pm 0.10$ (RMA regression, with $z_\text{LS}=1.16\pm 0.12$; see Ref.~\onlinecite{cav23}). 

We note that the static exponent $\nu$ (correlation length) and $\gamma$ (susceptibility) in the experiments are between the $\nu$ exponents of the 2D and 3D simulations. The dynamical exponent $z$ for the 2D and 3D is about 1, which is clearly different from the calculated from experimental data using RMA regression $z_\text{RMA}=1.37$ (using LS regression, $z_\text{LS}=1.16$ is still larger than in the 2D or 3D simulations). No data about $\varphi$ exist at the present time.

Data from natural swarms include observations under different environmental conditions and different number of insects or even midges of different species are involved. If we believe that swarms live in the criticality region of the scale-free-chaos phase transition, data from natural swarms will correspond to points on the critical curves of parameter space that have different values of $\eta$ and $N$. Recently, we have tried to mimic data from experimental observations by using a mixture of data from numerical simulations of the 3D HCVM that have different values of $N$ and $\eta$ on the critical curves $\beta_c$ and $\beta_0$ \cite{gon23arxiv}. While we get the same values of the exponent $z$ using LS or RMA regression if we simulate a single value of $\beta_c$ or $\beta_0$, the LS and RMA values of $z$ are different for the mixture of data. The resulting exponents of the 3D HCVM are $\nu=0.43\pm 0.03$, $\gamma=0.92\pm 0.13$, $z=1.37\pm 0.10$ (calculated using RMA regression, with LS regression we get $z=1.24\pm 0.11$), $\varphi=z\nu$ \cite{gon23arxiv}. These values are very close to the experimental ones listed above. Contrastingly, the mixtures of data that produce Fig.~\ref{fig13} for the 2D HCVM yield critical exponents $\nu=0.34\pm 0.05$, $z_\text{LS}\approx z_\text{RMA}\approx 1.2$: the static exponent $\nu$ is within the range of the experimental one but the dynamic exponent is somewhat smaller.

The theories based on the ordering phase transition predict accurately the dynamical critical exponent ($z=1.35$ for the active version of models E/F and G in Ref.~\cite{hoh77}), but fail to predict the static critical exponents. In Ref.~\cite{cav23} in addition to $z=1.35$, $\nu=0.748$, $\gamma=1.171$ are obtained, when the observed values are $\nu=0.35\pm 0.10$ and $\gamma=0.9\pm 0.2$ \cite{att14}. Furthermore, these theories fail to predict the limited collapse of the NDCCF \cite{cav17}, or the shape of the swarm \cite{att14,sin17}. 

In conclusion, we have analyzed the scale-free-chaos phase transition of the 2D HCVM based on numerical simulations of values of noise and number of particles on the critical curve $\beta_c$ separating single from multicluster chaotic attractors. The shape of the swarm (condensed core plus a vapor of particles entering and leaving it) and the partial collapse of the NDCCF in terms of rescaled time are the same as both 3D simulations and experimental data. The value of the static critical exponents $\nu$ and $\gamma$ are close to those obtained from simulations of the 3D HCVM and from experimental data. The dynamical exponent $z$ is different from that of the 3D HDVM and from experiments. We could not investigate the critical curve $\beta_0(N;\eta)$ separating chaotic and non-chaotic attractors due to non-convergence of the Benettin algorithm used to calculate the LLE at values of the confinement parameter that are too small. It would be interesting to study the HCVM having different confinement parameters in the vertical and transversal directions. This would be closer the observations of natural swarms, which are elongated in the vertical direction, and the results might be interpolations between the 2D and 3D models. These interpolations could also be useful to study renormalization group properties of the scale-free chaos phase transition based on the quasiperiodic route to chaos \cite{cen10,ber87}.

{\bf Acknowledgments.} This work has been supported by the FEDER/Ministerio de Ciencia, Innovaci\'on y Universidades -- Agencia Estatal de Investigaci\'on grants PID2020-112796RB-C21 (RGA) and PID2020-112796RB-C22 (LLB), by the Madrid Government (Comunidad de Madrid-Spain) under the Multiannual Agreement with UC3M in the line of Excellence of University Professors (EPUC3M23), and in the context of the V PRICIT (Regional Programme of Research and Technological Innovation). RGA acknowledges support from the Ministerio de Econom\'\i a y Competitividad of Spain through the  Formaci\'on de Doctores program Grant PRE2018-083807 cofinanced by the European Social Fund.


\begin{thebibliography}{}
\bibitem{pus22}A. Puscian, E. Knapska, Blueprints for measuring natural behavior. Iscience {\bf 25}, 104635 (2022).
\bibitem{bal08} M. Ballerini, N. Cabibbo, R. Candelier, A. Cavagna, E. Cisbani, I. Giardina, A. Orlandi, G. Parisi, A. Procaccini, M. Viale, V. Zdravkovic, Empirical investigation of starling flocks: a benchmark study in collective animal behaviour. Animal Behaviour {\bf 76}, 201-215 (2008).
\bibitem{gin15} F. Ginelli, F. Peruani, M.-H. Pillot, H. Chat\'e, G. Theraulaz, R. Bon, Intermittent collective dynamics emerge from conflicting imperatives in sheep herds. PNAS {\bf 112}(41), 12729-12734 (2015).
\bibitem{her11} J. E. Herbert-Read, A. Perna, R. P. Mann, T. M. Schaerf, D. J. T. Sumpter, A. J. W. Ward, Inferring the rules of interaction of shoaling fish. Proc. Natl Acad. Sci. USA {\bf 108}, 18726-18731 (2011).
\bibitem{buh06} J. Buhl, D. J. T. Sumpter, I. D. Couzin, J. J. Hale, E. Despland, E. R. Miller, S. J. Simpson, From disorder to order in marching locusts. Science {\bf 312}, 1402-1406 (2006).
\bibitem{att14plos} A. Attanasi, A. Cavagna, L. Del Castello, I. Giardina, S. Melillo, L. Parisi, O. Pohl, B. Rossaro, E. Shen, E. Silvestri, M. Viale, Collective behaviour without collective order in wild swarms of midges. PLoS Comput. Biol. {\bf 10}(7), e1003697 (2014).
\bibitem{kon19} K. Kondrakiewicz, M. Kostecki, W. Szadzinska, E. Knapska, Ecological validity of social interaction tests in rats and mice. Genes, Brain and Behavior {\bf 18}, e12525 (2019).
\bibitem{che23} X. Chen, M. Winiarski, A. Puscian, E. Knapska, T. Mora, A. M. Walczak, Modelling collective behavior in groups of mice housed under semi-naturalistic conditions. bioRxiv https://doi.org/10.1101/2023.07.26.550619
\bibitem{mor11}T. Mora, W. Bialek, Are biological systems poised at criticality?  J. Stat. Phys. {\bf 144}, 268-302 (2011).
\bibitem{bia12} W. S. Bialek, Biophysics: Searching for Principles (Princeton University Press, Princeton, 2012).
\bibitem{hua87}K. Huang, {\em Statistical Mechanics. 2nd ed} (Wiley, NY, 1987).
\bibitem{wil83} K. G. Wilson, The renormalization group and critical phenomena. Rev. Mod. Phys. {\bf 55}, 583-600 (1983).
\bibitem{ami05}D. J. Amit, V. Martin-Mayor, {\em Field Theory, The Renormalization Group and Critical Phenomena, 3rd ed} (World Scientific, Singapore, 2005).
\bibitem{hoh77}P. C. Hohenberg, B. I. Halperin, Theory of dynamic critical phenomena. Rev. Mod. Phys. {\bf 49}, 435-479 (1977).
\bibitem{cav18} A. Cavagna, I. Giardina, T.S. Grigera, The physics of flocking: Correlation as a compass from experiments to theory. Phys. Rep. {\bf 728}, 1-62 (2018).
\bibitem{vic95}T. Vicsek, A. Czir\'ok, E. Ben-Jacob, I. Cohen, O. Shochet, Novel type of phase transition in a system of self-driven particles. Phys. Rev. Lett. {\bf 75}, 1226-1229 (1995).
\bibitem{vic12}T. Vicsek, A. Zafeiris, Collective motion. Phys. Rep. {\bf 517}, 71-140 (2012).
\bibitem{cha20} H. Chat\'e, Dry aligning dilute active matter. Ann. Rev. Cond. Matter Phys. {\bf 11}, 189-212 (2020).
\bibitem{oku86}A. Okubo, Dynamical aspects of animal grouping: Swarms, schools, flocks, and herds. Adv. Biophys. {\bf 22}, 1-94 (1986).
\bibitem{kel13} D. H. Kelley, N. T. Ouellette, Emergent dynamics of laboratory insect swarms. Sci. Rep. {\bf 3}, 1073 (2013).
\bibitem{gor16} D. Gorbonos, R. Ianconescu, J. G. Puckett, R. Ni, N. T. Ouellette, N. S. Gov, Long-range acoustic interactions in insect swarms: an adaptive gravity model. New J. Phys. {\bf 18}, 073042 (2016).
\bibitem{cav17} A. Cavagna, D. Conti, C. Creato, L. Del Castello, I. Giardina, T. S. Grigera, S. Melillo, L. Parisi, M. Viale, Dynamic scaling in natural swarms. Nat. Phys. {\bf 13}, 914-918 (2017).
\bibitem{gon23}R. Gonz\'alez-Albaladejo, A. Carpio, L. L. Bonilla, Scale free chaos in the confined Vicsek flocking model. Phys. Rev. E {\bf 107}, 014209 (2023). 
\bibitem{gon23mf}R. Gonz\'alez-Albaladejo, L. L. Bonilla, Mean-field theory of chaotic insect swarms. Phys. Rev. E {\bf 107}, L062601 (2023). 
\bibitem{gon23arxiv}R. Gonz\'alez-Albaladejo, L. L. Bonilla, Power laws of natural swarms are fingerprints of an extended critical region. arXiv:2309.05064 
\bibitem{att14} A. Attanasi, A. Cavagna, L. Del Castello, I. Giardina, S. Melillo, L. Parisi, O. Pohl, B. Rossaro, E. Shen, E. Silvestri, M. Viale, Finite-size scaling as a way to probe near-criticality in natural swarms. Phys. Rev. Lett. {\bf 113}, 238102 (2014).
\bibitem{cav23} A. Cavagna, L. Di Carlo, I. Giardina, T. S. Grigera, S. Melillo, L. Parisi, G. Pisegna, M. Scandolo, Natural swarms in 3.99 dimensions. Nat. Phys. {\bf 19}, 1043-1049 (2023). 
\bibitem{sin17} M. Sinhuber, N. T. Ouellette, Phase coexistence in insect swarms. Phys. Rev. Lett. {\bf 119}, 178003 (2017).
\bibitem{dow55}J. A. Downes, Observations on the swarming flight and mating of Culicoides (Diptera: Ceratopogonidae). Trans. R. Entomol. Soc. London {\bf 106},  213-236 (1955).
\bibitem{lew17}J. M. Lewis, M. S. Turner, Density distributions and depth in flocks. J. Phys. D: Appl. Phys. {\bf 50}, 494003 (2017).
\bibitem{roy19}S. Roy, M. J. Shirazi, B. Jantzen, N. Abaid, Effect of visual and auditory sensing cues on collective behavior in Vicsek models. Phys. Rev. E {\bf 100}, 062415 (2019).
\bibitem{gei22}D. Geiss, K. Kroy, V. Holubec, Signal propagation and linear response in the delay Vicsek model. Phys. Rev. E {\bf 106}, 054612 (2022).
\bibitem{kum21} V. Kumar, R. De, Efficient flocking: metric versus topological interactions. R. Soc. Open  Sci. {\bf 8}, 202158 (2021).
\bibitem{rey17}A. M. Reynolds, M. Sinhuber, N. T. Ouellette, Are midge swarms bound together by an effective velocity-dependent gravity? Eur. Phys. J. E {\bf 40}, 46 (2017). 
\bibitem{ben80} G. Benettin, M. Casartelli, L. Galgani, A. Giorgilli, J. M. Strelcyn, Lyapunov characteristic exponents for smooth dynamical systems and for hamiltonian systems; A method for computing all of them. Part 2: Numerical application. Meccanica {\bf 15}, 21-30 (1980).
\bibitem{gao06} J. B. Gao, J. Hu, W. W. Tung, Y. H. Cao, Distinguishing chaos from noise by scale-dependent Lyapunov exponent. Phys. Rev. E {\bf 74}, 066204 (2006).
\bibitem{ott93} E. Ott, {\em Chaos in dynamical systems} (Cambridge University Press, Cambridge UK 1993).
\bibitem{gao94} J. B. Gao, Z. M. Zheng, Direct dynamical test for deterministic chaos and optimal embedding of a chaotic time series. Phys. Rev. E {\bf 49}, 3807-3814 (1994).
\bibitem{cen10} M. Cencini, F. Cecconi, A. Vulpiani, {\em Chaos. From simple models to complex systems} (World Scientific, New Jersey 2010). 
\bibitem{gol85}M. Golubitsky, D. G. Schaeffer, {\em Singularities and Groups in Bifurcation Theory, vol I} (Springer-Verlag, New York 1985).
\bibitem{ber87}P. Berg\'e, Y. Pomeau, and G. Vidal, G. {\em Order within Chaos: Towards a Deterministic Approach to Turbulence, 2nd ed} (John Wiley \& Sons: Toronto, Canada, 1987).
\end{thebibliography}
\end{document}